\documentstyle[12pt] {article}
\newcommand{\beq}{\begin{equation}}
\newcommand{\eeq}{\end{equation}}
\newcommand{\beqa}{\begin{eqnarray}}
\newcommand{\eeqa}{\end{eqnarray}}
\newcommand{\ba}{\begin{array}}
\newcommand{\ea}{\end{array}}

\begin{document}

\begin{center}
{\large \bf Quantal Overlapping Resonance Criterion: \\
the Pullen Edmonds Model}
\footnote{This work has been partially supported by the Ministero \\
dell'Universit\`a e della Ricerca Scientifica e Tecnologica (MURST).}
\end{center}

\vspace{0.5 cm}

\begin{center}
{\bf S. Graffi}  \\
Dipartimento di Matematica dell'Universit\`a
di Bologna, \\
Piazza di Porta S. Donato 5, I 40127 Bologna, Italy\\
\end{center}

\vskip 0.5 truecm

\begin{center}
{\bf V.R. Manfredi}
\footnote{Author to whom all correspondence and reprint requests should be
addressed. \\E--Mail: VAXFPD::MANFREDI, MANFREDI@PADOVA.INFN.IT} \\
Dipartimento di Fisica ``G. Galilei" dell'Universit\`a
di Padova, \\
INFN, Sezione di Padova, \\
Via Marzolo 8, I 35131 Padova, Italy $^{(+)}$ \\
\end{center}

\vskip 0.5 truecm

\begin{center}
{\bf L. Salasnich}  \\
Dipartimento di Fisica dell'Universit\`a di Firenze, \\
INFN, Sezione di Firenze, \\
Largo E. Fermi 2, I 50125 Firenze, Italy \\
\end{center}

\vskip 0.5 truecm

\begin{center}
\end{center}

\newpage

{\begin{center}
{\bf Abstract}
\end{center}}
\vskip 0.5 truecm
\par
In order to highlight the onset of chaos in the Pullen-Edmonds model
a quantal analog of the resonance overlap criterion has been examined.
A quite good agreement between analytical and numerical results is obtained.

\newpage

\par
Recently great interest has been shown in the so--called {\it quantum
chaos}, or more precisely {\it quantum chaology}, i.e. the study
of quantal systems which are chaotic in the classical limit
$\hbar \to 0$ [1].
\par
The aim of this work is to apply a quantal analog of the resonance
criterion [2] to the Pullen Edmonds model [3], in order to highlight the onset
of chaos in a schematic model.
A requisite for the occurrence of quantum chaos
is that the spectrum behaves in an irregular way as a function of
the coupling constant: in particular if multiple avoided crossings occur
(see for example [4] and references quoted therein).
In fact, as stressed by Berry [5], in {\it quasi--integrable systems} the
presence of many quasi--crossings leads to a chaotic behaviour.
Using the semiclassical quantization we calculate the critical value
$\chi_{c}$ of the coupling constant corresponding to the
intersection of two neighboring quantal separatrices.
This criterion is, in some sense, the quantal
counterpart of the method of overlapping resonances developed by
Chirikov [6]. We point out however that the Pullen Edmonds hamiltonian is
the perturbation of an intrinsically resonant hamiltonian
(see e.g.[7] and references quoted therein) which yields intrinsically
degenerate quantum unperturbed levels:
hence we first have have to extend to this situation the argument of [2],
valid in the case of accidental degenaracy.
\par
As discussed by Pullen and Edmonds [4], the hamiltonian:
\beq
H={1\over 2}(p_1^2+p_2^2)+{1\over 2}(q_1^2+q_2^2)+
       \chi q_1^2 q_2^2,
\eeq
exhibits a transition from order to chaos as a function of the coupling
constant $\chi$. We introduce the new variables $(I,\theta )$
by the canonical transformation:
\beq
\left\{
\ba{ccc}
q_{i}&=&\sqrt{2I_{i}}\cos{\theta_{i}} \\
p_{i}&=&\sqrt{2I_{i}}\sin{\theta_{i}}.
\ea
\right.
\;\;\; i=1,2.
\eeq
Then (1) becomes:
\beq
H=I_{1}+I_{2}+4\chi I_{1}I_{2}
       \cos^{2}{\theta_{1}}\cos^{2}{\theta_{2}}.
\eeq
By the new canonical transformation:
\beq
\left\{
\ba{ccc}
A_{1}&=&I_{1}+I_{2} \\
A_{2}&=&I_{1}-I_{2}
\ea
\right.
\;\;\;\;\;
\left\{
\ba{ccc}
\theta_{1}&=&\phi_{1}+\phi_{2} \\
\theta_{2}&=&\phi_{1}-\phi_{2},
\ea
\right.
\eeq
$H$ can be written:
\beq
H=A_{1}+\chi (A_{1}^{2}-A_{2}^{2})
       \cos^{2}{(\phi_{1}+\phi_{2})}\cos^{2}{(\phi_{1}-\phi_{2})}.
\eeq
We now eliminate the dependence on the angles to order $\chi^2$ by
resonant (or secular) canonical perturbation theory [8]. First we average
on the fast variable ($\phi_{1}$). This yields:
\beq {1\over
2\pi}\int_{0}^{2\pi}d\phi_{1}
\cos^{2}{(\phi_{1}+\phi_{2})}\cos^{2}{(\phi_{1}-\phi_{2})} ={1\over
8}(2+\cos{4\phi_2}),
\eeq
and:
\beq
{\bar H}_{cl}=A_{1}+{\chi \over 8}(A_{1}^{2}-A_{2}^{2})
(2+\cos{4\phi_2}).
\eeq
The dependence on $\phi_2$ is now eliminated by a second canonical
transformation. The Hamilton--Jacobi equation for the perturbation part
is indeed:
\beqa [A_{1}^{2}-({\partial S\over \partial
\phi_{2}})^{2}] (2+\cos{4\phi_{2}})=K,
\\
{\partial S\over \partial \phi_{2}}=\pm
\sqrt{ A_{1}^{2}(2+\cos{4\phi_{2}})-K\over 2+\cos{4\phi_{2}} }.
\eeqa
and thus the Hamiltonian (7) becomes:
\beq
{\bar H}=B_{1}+{\chi \over 8}K(B_{1},B_{2}),
\eeq
where:
\beq
B_{1}=A_{1}, \;\;\;\;
B_{2}={1\over 2\pi}\oint d\phi_{2} {\partial S\over \partial \phi_{2}}.
\eeq
It appears from the structure of equation (8) that the motion of our
system is similar that of a simple pendulum:
\beq
\ba{cc}
0<K<B_{1}^{2} \;\;\;\; & rotational \; motion
\\
K=B_{1}^{2}   \;\;\;\; & separatrix
\\
B_{1}^{2}<K<3B_{1}^{2} \;\;\;\; & librational \; motion.
\ea
\eeq
On the separatrix:
\beq
B_{1}^{2}(2+\cos{4\phi_{2}})=K,
\eeq
and:
\beq
B_{2}=\pm {2\over \pi}\int_{a}^{b}dx
\sqrt{ B_{1}^{2}(2+\cos{4x})-K\over 2+\cos{4x} },
\eeq
where:
\beq
\ba{cc}
a=-{\pi \over 4}, \;\; b={\pi\over 4} \;\;\;\; & rotational \; motion
\\
a=\phi_{-}(K,B_{1}), \;\; b=\phi_{+}(K,B_{1})
\;\;\;\; & librational \; motion
\ea
\eeq
with:
\beq
\phi_{\pm}(K,B_{1})=\pm {1\over 4}\arccos ({K\over B_1^2}-2).
\eeq
\par
The semiclassical quantization on the Hamiltonian (10) may be used. Set:
\beq
B_{1}=m_{1}\hbar , \;\; B_{2}=m_{2}\hbar ;
\eeq
then [9], up to terms of order $\hbar$, the quantum spectrum is:
\beq
E_{m_{1},m_{2}}=m_{1}\hbar +{\chi \over 8}K(m_{1}\hbar , m_{2}\hbar ),
\eeq
where $K$ is implicitly defined by the relation:
\beq
m_{2}\hbar =\pm {2\over \pi}\int_{a}^{b}dx
\sqrt{ (m_{1}\hbar )^{2}(2+\cos{4x})-K\over 2+\cos{4x} },
\eeq
and:
\beq
\ba{cc}
a=-{\pi\over 4}, \;\; b={\pi\over 4} \;\;\;\; & 0<K<(m_{1}\hbar )^{2}
\\
a=\phi_{-}(K,B_{1}), \;\; b=\phi_{+}(K,B_{1})
\;\;\;\; & (m_{1}\hbar )^{2}<K<3(m_{1}\hbar )^{2}
\ea
\eeq
\par
On the separatrix, where $K=(m_1\hbar )^2$, $m_2 =\pm \alpha m_1$,
with:
\beq
\alpha ={2\over \pi}\int_{-{\pi\over 4}}^{\pi\over 4}
dx \sqrt{1+\cos{4x}\over 2+\cos{4x}}.
\eeq
Figure 1 shows $K$ {\it vs} $m_{2}$.
\par
By (18) it is not difficult to calculate the critical value
$\chi_{c}$ of the coupling constant corresponding to the
intersection of the separatrices of two neighboring multiplets:
\beq
(m_1+1)\hbar +{\chi \over 8}K[(m_1+1)\hbar ,\alpha (m_1+1)\hbar ]=
m_1\hbar +{\chi \over 8}K(m_1\hbar ,\alpha m_1\hbar ),
\eeq
and so:
\beq
\chi_{c}={ -8\hbar \over
K[(m_1+1)\hbar ,\alpha (m_1+1)\hbar ]-K(m_1\hbar ,\alpha m_1\hbar ) }.
\eeq
The denominator of (23) can be evaluated by the Taylor expansion and
finally:
\beq
\chi_{c}= \left[ { -8 \over
{\partial K\over \partial B_{1}}-
\alpha{\partial K \over \partial B_{2}} }
\right]_{B_1=m_1\hbar ,B_2=\alpha m_2\hbar}.
\eeq
K is implicitly defined by the relation:
\beq
F[B_1,B_2,K(B_1,B_2)]=B_2-{\pi\over 2}
\int_{-{\pi\over 4}}^{\pi\over 4}dx
\sqrt{ B_1^{2}(2+\cos{4x})-K\over 2+\cos{4x} }=0,
\eeq
or:
\beq
F(B_{1},B_{2},K)=B_{2}-\Phi (B_{1},K)=0.
\eeq
As a function of $\Phi$, $\chi_{c}$ can be written:
\beq
\chi_{c}= \lim_{K\to B_1^2} \left[ { 8{\partial \Phi\over \partial K}
\over
\alpha -{\partial \Phi\over \partial B_{1}}}
\right]_{B_1=m_1\hbar},
\eeq
where:
$$
{\partial \Phi\over \partial K}=-{1\over \pi}
\int_{-{\pi\over 4}}^{\pi\over 4} dx
{ 1\over \sqrt{(2+\cos{4x})[B_1^2(2+\cos{4x})-K]} }
$$
\beq
{\partial \Phi\over \partial B_1}={2\over \pi}
\int_{-{\pi\over 4}}^{\pi\over 4}dx
\sqrt{ {B_1^2(2+\cos{4x})\over B_1^2(2+\cos{4x})-K} }.
\eeq
It is not difficult to show that:
\beq
\chi_c={4\over m_1\hbar}.
\eeq
Figure 2 shows $\chi_{c}$ {\it vs} $m_1\hbar$. If we take the energy
$E\simeq m_1\hbar$ we obtain $\chi_c \simeq 4/ E$,
a quantal rule for the onset of classical chaos.
\par
The transition order--chaos in systems with two
degrees of freedom may be also studied by the curvature criterion of
potential energy [10,11]. It is however important to point out that
{\it in general} the curvature
criterion guarantees only a {\it local instability} and should therefore
be combined with the Poincar\`e sections [14]. For a fuller discussion of
this point see [12].
\par
At low energy, the motion near the minimum of the potential
$V(q_1,q_2)={1\over 2} (q_1^2+q_2^2)+\chi q_1^2 q_2^2$,
where the curvature is positive, is periodic or quasiperiodic and is
separated from the region of instability by a line of zero curvature;
if the energy is increased, the system will be for some initial conditions
in a region of negative curvature, where the motion is chaotic.
In accordance with this scenario, the energy of order$\to$chaos transition
$E_c$ is equal to the minimum value of the line of zero gaussian
curvature $K(q_1 ,q_2 )$ of the potential--energy surface of the system.
For our potential the gaussian curvature vanishes at the points
that satisfy the equation:
\beq
{\partial^2 V \over \partial q_1^2}
{\partial^2 V \over \partial q_2^2}-
({\partial^2 V \over \partial q_1 \partial q_2})^2=
(2\chi q_1^2 +1)(2\chi q_2^2 +1) -16 \chi^2 q_1^2 q_2^2 =0.
\eeq
It is easy to show that the minimal energy on the
zero--curvature line is given by [13]:
\beq
E_c=V_{min}(K=0,\bar{r_1})={4\over 3\chi},
\eeq
and occurs at $\bar{q_1}=\pm \sqrt{1\over 2\chi}$. This result is in
rough agreement with equation (29) and with the
numerical calculations of Poincar\`e sections [14] obtained by
authors of reference [3]. The discrepancy between (29) and (31)
is due to the fact that the curvature criterion describes the onset of
{\it local chaos} associated to
quasi--crossings between levels of the same multiplet (see also
reference [3]) meanwhile the quantum resonance criterion describes
the onset of {\it widespread chaos} associated to quasi--crossings
between separatices of different multiplets.

\begin{center}
*****
\end{center}

The authors are greatly indebted to Dr. Stefano Isola for many useful
discussions.

\newpage

\section*{Figure Captions}
\vspace{0.6 cm}
\parindent=0.pt

Figure 1: The semiclassical energy $K$ {\it vs} $m_2$ for a fixed $m_1$:
$0<K<1$ rotational motion (a) and $1<K<3$ librational motion (b).

Figure 2: The critical value $\chi_c$ {\it vs} $m_1\hbar$.

\newpage

\section*{References}
\vspace{0.6 cm}
\parindent=0.pt

[1] A.M. Ozorio de Almeida: {\it Hamiltonian Systems: Chaos and
Quantization}, Cambridge University Press (1990);
M.C. Gutzwiller: {\it Chaos in Classical and Quantum Mechanics},
Berlin, Heidelberg, New York: Springer Verlag (1990);
{\it From Classical to Quantum Chaos}, Conference Proceedings SIF,
vol 41, Ed. G.F. Dell'Antonio, S. Fantoni, V.R. Manfredi (Editrice
Compositori, Bologna, 1993).

[2] S. Graffi, T. Paul, H.J. Silverstone: Phys. Rev. Lett. {\bf 59}, 255
(1987); S. Graffi, T. Paul, H.J. Silverstone: Phys. Rev. A {\bf 37},
2214 (1988)

[3] R.A. Pullen, R.A. Edmonds: J. Phys. A {\bf 14}, L477
(1981)

[4] V.R. Manfredi, L. Salasnich, L. Dematt\`e: {\it Phys. Rev.} E
{\bf 47}, 4556 (1993)

[5] M.V. Berry: in {\it Chaootic Behaviour in Quantum Systems},
Nato Series B, vol. 120, Ed. G. Casati (1983)

[6] B.V. Chirikov: {\it Phys. Rep.} {\bf 52}, 263 (1979)

[7] S. Graffi: in {\it Probabilistic Methods in Mathematical Physics},
Ed. F. Guerra, M.I. Loffredo, C. Marchioro (World Scientific, New York,
1991)

[8] M. Born: {\it Mechanics of the Atom} (Bell, London, 1960); T. Uzer,
D.W. Noid, R.A. Marcus: {\it J. Chem. Phys.} {\bf 79}, 4412 (1983)

[9] S. Graffi, V.R. Manfredi, L. Salasnich: Preprint DFPD/TH/93, to be
published.

[10] M. Toda: {\it Phys. Lett.} A {\bf 48}, 335 (1974)

[11] Yu. Bolotin, V.Yu. Gonchar, E.V. Inopin, V.V. Levenko, V.N. Tarasov,
N.A. Chekanov: {\it Sov. I. Part. Nucl.} {\bf 20}, 372 (1989)

[12] G. Benettin, R. Brambilla, L. Galgani: {\it Physica} A {\bf 87},
381 (1977)

[13] W.H. Steeb, A. Kunick: {\it Lett. Nuovo Cimento} {\bf 42}, 89
(1985)

[14] H. Poincar\`e: {\it New Methods of Celestial Mechanics}, vol. 3, ch.
27 (Transl. NASA Washington DC 1967);
M. Henon: {\it Physica} D {\bf 5}, 412 (1982)

\end{document}